\documentclass[aps,pra,reprint,amsmath,amssymb,superscriptaddress,showpacs,nobibnotes,twocolumn]{revtex4-2}

\usepackage{graphicx}
\usepackage{dcolumn}
\usepackage{bm}
\usepackage{amsmath}
\usepackage{amssymb,amsthm}
\usepackage{epsfig,color}
\usepackage[sans]{dsfont}
\usepackage{romannum}
\usepackage{upgreek}
\usepackage{siunitx}

\definecolor{nblue}{rgb}{0.2,0.2,0.7}
\definecolor{ngreen}{rgb}{0.2,0.6,0.2} 
\definecolor{nred}{rgb}{0.7,0.2,0.2}
\definecolor{nblack}{rgb}{0,0,0}

\newcommand{\ket}[1]{|#1\rangle}
\newcommand{\ketbra}[1]{\ket{#1}\!\bra{#1}}
\newcommand{\bra}[1]{\langle#1|}
\newcommand{\braket}[2]{\langle #1 | #2 \rangle}
\newcommand{\proj}[1]{\ket{#1}\!\bra{#1}}
\newcommand{\tr}{\text{tr}}

\renewcommand{\H}{\mathcal{H}}

\def\g{\mathrm{guess}}
\def\E{\mathrm{E}}
\def\N{\mathcal{N}}

\def\tr{\mbox{tr}}
\def\bea{\begin{eqnarray}}
\def\eea{\end{eqnarray}}

\begin{document}



\title{ Single-Copy Certification of Two-Qubit Gates without Entanglement}


\author{Yujun Choi}
\affiliation{Center for Quantum Information, Korea Institute of Science and Technology (KIST), Seoul, 02792, Republic of Korea}
\affiliation{Department of Physics, Yonsei University, Seoul, 03722, Republic of Korea}

\author{Tanmay Singal}
\affiliation{Institute of Physics, Faculty of Physics, Astronomy, and Informatics, Nicolaus Copernicus University, Grudziadzka 5, 87-100 Torun, Poland}


\author{Young-Wook Cho}
\affiliation{Center for Quantum Information, Korea Institute of Science and Technology (KIST), Seoul, 02792, Republic of Korea}

\author{Sang-Wook Han}
\affiliation{Center for Quantum Information, Korea Institute of Science and Technology (KIST), Seoul, 02792, Republic of Korea}
\affiliation{Division of Nano \& Information Technology, KIST School, Korea University of Science and Technology, Seoul 02792, Republic of Korea}

\author{Kyunghwan Oh}
\affiliation{Department of Physics, Yonsei University, Seoul, 03722, Republic of Korea}

\author{Sung Moon}
\affiliation{Center for Quantum Information, Korea Institute of Science and Technology (KIST), Seoul, 02792, Republic of Korea}
\affiliation{Division of Nano \& Information Technology, KIST School, Korea University of Science and Technology, Seoul 02792, Republic of Korea}
\author{Yong-Su Kim}
\email{yong-su.kim@kist.re.kr}
\affiliation{Center for Quantum Information, Korea Institute of Science and Technology (KIST), Seoul, 02792, Republic of Korea}
\affiliation{Division of Nano \& Information Technology, KIST School, Korea University of Science and Technology, Seoul 02792, Republic of Korea}

\author{Joonwoo Bae }
\email{joonwoo.bae@kaist.ac.kr}
\affiliation{ School of Electrical Engineering, Korea Advanced Institute of Science and Technology (KAIST), 291
Daehak-ro, Yuseong-gu, Daejeon 34141, Republic of Korea}
 
\date{\today}

\begin{abstract}

A quantum state transformation can be generally approximated by single- and two-qubit gates. This, however, does not hold with noisy intermediate-scale quantum technologies due to the errors appearing in the gate operations, where errors of two-qubit gates such as controlled-NOT and SWAP operations are dominated. In this work, we present a cost efficient single-copy certification for a realization of a two-qubit gate in the presence of depolarization noise, where it is aimed to identify if the realization is noise-free, or not. It is shown that entangled resources such as entangled states and a joint measurement are not necessary for the purpose, i.e., a noise-free two-qubit gate is not needed to certify an implementation of a two-qubit gate. A proof-of-principle demonstration is presented with photonic qubits. 
\end{abstract}


\maketitle 

\section{Introduction}

All quantum operations realized within the state of the art technologies, dubbed noisy intermediate-scale quantum (NISQ) devices, contain noise \cite{Preskill2018quantumcomputingin}. Two-qubit gates are of particular importance as they can generate entangled states that are a general resource for quantum information processing \cite{Kitaev_1997, PhysRevA.52.3457} and, at the same time, can also play a role of propagating local errors over a quantum circuit for designed information processing. This applies to two-qubit gates in general, almost all of which can be used to construct a set of universal gates \cite{PhysRevLett.75.346}. In fact, the error rates of two-qubit gates with NISQ devices are significantly higher than single-qubit ones, see e.g. \cite{PhysRevLett.121.220502, PhysRevA.100.052315}. It is clear that quantum advantages cannot be achieved when local errors are high enough and accumulated in a quantum circuit. 
  
The realization of two-qubit gates with a low error rate is thus identified as one of the key building blocks towards quantum advantages with NISQ technologies in practice. Schemes for mitigating quantum errors developed recently can be applied, e.g., \cite{Songeaaw5686, PhysRevX.8.031027, PhysRevLett.119.180509, mitigatingreadout}, where single-qubit gates having a relatively much lower error rate are complemented to systematically suppress the local errors. Or, it may be attempted to detect a noisy two-qubit operation beforehand so that it is to be replaced with a cleaner one having a lower error rate. In both cases, it is essential to efficiently identify a noisy implementation of a two-qubit gate placed in a quantum circuit. 

In a verification of a quantum operation, a measurement should be applied repeatedly in order to single out a unique operation from measurement data . On the other extreme, a conclusion from a single-shot measurement is, however, correct with some probability at its best. The success probability is limited by fundamental principles \cite{doi:10.1063/1.3298647, PhysRevLett.107.170403}. It is therefore required to optimize quantum resources, such as states, operations and measurements, in order to maximize the probability of making a correct conclusion. The advantage lies at the fact that a single-shot measurement is cost efficient. For instance, there are multipartite entangled states that can be certified by a single-shot measurement with a probability, which in fact converges to the certainty as the number of parties increases \cite{Dimic:2018aa}. 

In this work, we present a single-copy certification for a realization of a two-qubit gate in the presence of depolarization noise, so that the decision of replacing it with a cleaner one or placing local unitaries for mitigating errors can be effectively made for practical purposes. For a two-qubit gate $U$, its noisy counterpart containing depolarization noise is denoted by a set of two-qubit channels 
\bea 
[\N_{U}] := \{ \N_{U}^{p} =  (1-p ) U (\cdot) U^{\dagger}  + p D (\cdot)~ :~ \forall p \in (0,1] \} ~~~~\label{eq:1}
\eea
where the complete depolarization map is denoted by $D(\cdot) = \mathrm{I}/4$ and a noise fraction by $p\in[0,1]$. The goal is to certify a realization of a two-qubit gate by a single-shot measurement. Therefore, an optimal discrimination of a two-qubit gate $U$ from the set $[ \N_{U} ]$ can be found as a tool for a single-shot certification by maximizing the success probability. For practical purposes, in addition, we include the constraint that non-entangled resources only are applied in the certification. Otherwise, one has to assume a noise-free two-qubit gate for preparing an entangled state or a joint measurement for a certification of a realization of another two-qubit gate. Without the assumption, a higher level of confidence is achieved on a certification.

We here show that the aforementioned single-copy certification can be optimally achieved without an entangled resource {\it at all}. That is, an optimal discrimination between a two-qubit gate $U$ and its noisy counterpart $[\N_{U}]$, while a noise fraction is unknown, can be achieved by unentangled resources only. We present a scheme of an optimal one-shot certification for two-qubit gates. The scheme can also be used to estimate a noise fraction when a measurement is repeatedly applied. The result holds true for all two-qubit gates. For a controlled-NOT (CNOT) gate having a particular importance in the construction of a quantum circuit, a single-shot certification with local resources is experimentally demonstrated with photonic qubits. The experimental scheme can be straightforwardly extended to other physical systems.  


\section{ Theory  }

The strategy here for the single-shot certification of a two-qubit gate exploits optimal discrimination between quantum channels. Given two quantum channels and their single use, it is attempted to maximize the probability of making a correct guess on average. While a two-qubit gate $U$ is attempted, it may appear as a noisy gate $\N_{U}^p \in [\N_U]$ in Eq. (\ref{eq:1}) with some {\it a priori} probability $q$, where a noise fraction $p$ is unknown. Then, the realization remains in a noise-free case $U$ with a probability $1-q$. The goal of certification is to find if a realization of a two-qubit gate remains noise-free, or not, where it is aimed to maximize the probability of making a correct conclusion. As is mentioned above, the problem is approached by optimal discrimination of a two-qubit gate $U$ from the set $[ \N_{U} ]$ in Eq. (\ref{eq:1}) by a single-shot measurement. 

\subsection{Minimum-error channel discrimination}

We first show optimal channel discrimination with a general measurement. For the practical purpose that we minimize experimental resources, ancillary systems are not applied here. Two quantum operations $U$ and $\N_{U}^p$ for some $p$ can be optimally discriminated by applying an input state and a measurement on the resulting states. Let $\rho$ denote a two-qubit state and $\{ \Pi_1 , \Pi_2\}$ positive-operator-valued-measure (POVM) elements for a two-outcome measurement. A detection event on $\Pi_1$ concludes a two-qubit gate $U$ and a detection on $\Pi_2$ its noisy counterpart, respectively. The probability that the conclusion is correct is denoted by the guessing probability, maximized over a two-qubit state and a measurement as follows, 
\bea
p_{\g}  & = & \max_{\rho} \max_{\Pi_1, \Pi_2}~ (1-q)\tr[ U \rho U^{\dagger} \Pi_1]   +  q \tr[ \N_{U}^{p} (\rho) \Pi_2]  \nonumber \\ 
& = &  \frac{1}{2} + \frac{1}{2} \max_{\rho}  \|  (1-q) U \rho U^{\dagger} - q \N_{U}^p (\rho) \|_1 \label{eq:pg1}
\eea
where $\| \cdot \|_1$ denotes the $L_1$-norm. The second equality follows from the result in minimum-error state discrimination between two quantum states ~\cite{holevo1974, helstrom1976, yuen1975}, see also related reviews \cite{bergou2004, bergou2007, barnett2009, bae2015}. Note that it suffices to consider pure states in the optimization.  

The guessing probability is computed as follows,
\bea
p_{\g}  & = & \frac{1}{2} \left(1 +  \frac{3}{4}pq + | 1-2q + \frac{3}{4}pq |  \right). \label{eq:pg2}
\eea
For $1-2q + \frac{3}{4}pq <0$, we have $p_{\g} =q$. This corresponds to the case where no measurement is actually applied \cite{PhysRevA.68.012306}. The optimal strategy is to guess a noisy channel $\N_{U}^p$ all the time according the {\it a priori} probability, without a measurement. For $1-2q + \frac{3}{4}pq \geq 0$, both POVM elements are non-zero: one can find an optimal measurement contains POVM elements as follows,
\bea
\Pi_1 = U |\psi\rangle \langle \psi | U^{\dagger}~~\mathrm{and}~~  \Pi_2 =  \mathrm{I} - U |\psi\rangle \langle \psi | U^{\dagger}.\label{eq:two}
\eea
for an optimal two-qubit state $|\psi\rangle$. The guessing probability is given by $p_{\g} = 1-q +3pq/4$. 

It is worth to notice that an optimal measurement in Eq. (\ref{eq:two}) does not depend on a noise fraction $p$. This means that minimum-error discrimination between a gate $U$ and a noisy channel $\N_{U}^{p}$ can be equivalently applied to that of the gate and a collection of the channels $[\N_U]$. This makes it possible to apply minimum-error channel discrimination in the certification of two-qubit gates. 

 
\subsection{Optimal channel discrimination with LOCC}

In this subsection, we show that the guessing probability in Eq. (\ref{eq:pg2}) from optimal channel discrimination can be achieved with local operations and classical communication (LOCC) only, without entangled resources. In fact, we present an optimal separable measurement explicitly. The derivation consists of a few steps as follows.

\subsubsection*{ Channel discrimination with LOCC}

To formalize optimal channel discrimination with LOCC, we write the guessing probability constrained by LOCC as follows,
\bea
p_{\g}^{(\mathrm{LOCC})}  = \frac{1}{2} + \frac{1}{2} \max_{\rho\in \mathrm{SEP}}  \|  (1-q) ~U \rho U^{\dagger} - q ~\N_{U}^p (\rho) \|_{\mathrm{LOCC}} \nonumber, \label{eq:pglocc}
\eea
where the LOCC norm has been operationally defined as $\| X \|_{\mathrm{LOCC}} = \sup_{\mathcal{M}\in \mathrm{ LOCC}} \| \mathcal{M} (X) \|_1 $ and $\mathcal{M}$ denotes a set of POVMs or quantum instruments associated to LOCC \cite{Matthews:2009aa}. The guessing probability $p_{\g}^{(\mathrm{LOCC})}$ can be obtained by computing an LOCC norm when a separable state is applied to one of the quantum operations.  

Let us first consider a discrimination task constrained by an unentangled measurement, which computes an LOCC norm.
\\

{\bf Proposition}. Suppose that for bipartite states $\{q_i ,\rho_i \}_{i=1}^2$ an optimal discrimination is achieved by a measurement $\{ \Pi_i\}_{i=1}^2$. The optimal discrimination can be achieved by LOCC, i.e.,
\bea
\|q_1 \rho_1 - q_2 \rho_2 \|_{\mathrm{LOCC}} = \|q_1 \rho_1 - q_2 \rho_2 \|_1
\eea 
if and only if normalized POVM elements $\{ \widetilde{\Pi}_i \}_{i=1}^2$, i.e., $\widetilde{\Pi}_i= \Pi_i / \tr[\Pi_i]$ that can be interpreted as quantum states, can be perfectly discriminated by an LOCC protocol. \\

This shows that an LOCC protocol for perfectly distinguishing normalized POVM elements leads to an optimal discrimination of two bipartite states $\{q_i ,\rho_i \}_{i=1}^2$ for which the POVM elements construct an optimal measurement. In the discrimination task via an LOCC protocol, a conclusion $\widetilde{\Pi}_i$ with certainty finds the corresponding state $\rho_i$ optimally for $i=1,2$, i.e., with a maximal probability of making a correct guess. 

Before proceeding to the proof, we describe the feature of a general LOCC protocol on a shared state $\rho_{AB}$. Without loss of generality, we assume that Alice first begins a protocol, in which $\{K_{j}^{A} \}$ denote her Kraus operators, i.e. it holds that $\sum_{j} {K_{j}^{A}}^{ \dagger} K_{j}^{A} = \mathrm{I}_A$. Alice's local operation on a shared state is described by $\{ K_{j}^{A} \otimes \i^{B} \}$. Bob acknowledges Alice's measurement outcome, denoted by $k_1$, according to which he devises local operations described by Kraus operators $\{ L_{j | k_1 }^{B} \}$ such that $\sum_j {L_{j | k_1 }^{B}}^{\dagger} L_{j | k_1 }^{B} = \mathrm{I}_B$. Let $l_1$ be Bob's outcome in the first round, after which the resulting state is given by, up to normalization,
\bea
\rho_{AB} ~\mapsto ~ (\mathrm{I}\otimes L_{l_{1} | k_1}^{B} ) (K_{k_1}^{A} \otimes \mathrm{I}) ~\rho_{AB} ~(K_{k_1}^{A} \otimes \mathrm{I})^{\dagger} ( \mathrm{I} \otimes L_{l_{1} | k_1}^{B})^{\dagger}. \nonumber
\eea
Note that this happens with the following probability
\bea
p_1 = \tr[ (\mathrm{I} \otimes  {{L_{l_{1} | k_1}^{B}}^\dagger} L_{l_{1} | k_1}^{B}  ) ( {{K_{k_1}^{A}}^\dagger} K_{k_1}^{A} \otimes \mathrm{I} ) \rho_{AB}]. \nonumber
\eea
According to the outcomes $(k_1, l_1)$, Alice decides local operations to apply, denoted by $\{ K_{j | (k_1,l_1)}^{A} \}$, and obtains an outcome denoted by $k_2$, corresponding to which Bob performs local operations $\{ L_{j |k_2 (k_1,l_1)}^{B} \}$. Let $(k_2,l_2)$ denote the measurement outcome in the second round. After the $n$-th round, we write the outcomes as 
\bea
(\vec{k}_n ,\vec{l}_n) := (k_n,l_n)(k_{n-1}, l_{n-1})\cdots (k_1, l_1). \nonumber
\eea
One can assume that, without loss of generality, an LOCC protocol terminates on the Bob's side with finite $n$. 

Then, the $n$-th Kraus operators of Alice and Bob can be generally written as
\bea
K_{(\vec{k}_{n} ,\vec{l}_{n} ) }^{A} & = &   K_{k_n | (\vec{k}_{n-1} ,\vec{l}_{n-1}) }^{A} K_{k_{n-1} | (\vec{k}_{n-2} ,\vec{l}_{n-2}) }^{A}\cdots K_{k_1}^{A}    \nonumber \\
L_{(\vec{k}_{n} ,\vec{l}_{n} ) }^{B} & = & L^{B}_{l_{n} | k_{n} (\vec{k}_{n-1} ,\vec{l}_{n-1}) } L^{B}_{l_{n-1} | k_{n-1} (\vec{k}_{n-2} ,\vec{l}_{n-2}) }\cdots L_{l_1|k_1}^{B}  \nonumber 
\eea
In this way, the resulting Kraus operators of Alice and Bob $\{ E_{( \vec{k}_n ,\vec{l}_n  )}^{\mathrm{AB_{LOCC}}} \}_{(\vec{k}_n ,\vec{l}_n )}  $ of the $n$ rounds for measurement outcomes $(\vec{k}_n ,\vec{l}_n )$ are described by
\bea
E_{ ( \vec{k}_n ,\vec{l}_n  )}^{\mathrm{AB_{LOCC}}}  =   K_{ (\vec{k}_{n } ,\vec{l}_{n } ) }^{A} \otimes L^{B}_{  (\vec{k}_{n } ,\vec{l}_{n } ) } \label{eq:KLOCC}
\eea
such that $\sum_{ ( \vec{k}_n ,\vec{l}_n  ) }   {E_{ ( \vec{k}_n ,\vec{l}_n  )}^{\mathrm{AB_{LOCC}}} }^{\dagger} E_{ ( \vec{k}_n ,\vec{l}_n  )}^{\mathrm{AB_{LOCC}}} =\mathrm{I}_{AB}$. With this description of LOCC, the proof of the aforementioned theorem is presented below.

\begin{proof}
\begin{subequations}
\label{eq:thm1}

($\Leftarrow$) Suppose the two states $\widetilde{\Pi_1}$ and $\widetilde{\Pi_2}$ can be perfectly discriminated by some LOCC protocol. This means that for all sequences of outcomes of the LOCC protocol, $(\vec{k}_n,\vec{l}_n)$, one can conclusively rule out one of the two states being present. This implies that all sequences $\{(k_n,l_n)\}$ can be partitioned into two classes: $\{(\vec{s}_n,\vec{t}_n) \}$ and $\{(\vec{v}_n,\vec{w}_n)\}$  such that POVM elements corresponding to them $ {E_{ ( \vec{s}_n ,\vec{t}_n  )}^{\mathrm{AB_{LOCC}}} }^{\dagger} E_{ ( \vec{s}_n ,\vec{t}_n  )}^{\mathrm{AB_{LOCC}}} $ and $ {E_{ ( \vec{v}_n ,\vec{w}_n  )}^{\mathrm{AB_{LOCC}}} }^{\dagger} E_{ ( \vec{v}_n ,\vec{w}_n  )}^{\mathrm{AB_{LOCC}}}$ satisfy the following
\bea
&& \tr[\widetilde{\Pi_2} ~ {E_{ ( \vec{s}_n ,\vec{t}_n  )}^{\mathrm{AB_{LOCC}}} }^{\dagger} E_{ ( \vec{s}_n ,\vec{t}_n  )}^{\mathrm{AB_{LOCC}}} ] =0~~\mathrm{and}~~\nonumber \\
&& \tr[\widetilde{\Pi_1}  ~{E_{ ( \vec{v}_n ,\vec{w}_n  )}^{\mathrm{AB_{LOCC}}} }^{\dagger} E_{ ( \vec{v}_n ,\vec{w}_n  )}^{\mathrm{AB_{LOCC}}}] = 0. \nonumber 
\eea
Since POVM elements of the LOCC protocol is complete, we have that
\bea
&& \sum_{(\vec{s}_n,\vec{t}_n)}{E_{ ( \vec{s}_n ,\vec{t}_n  )}^{\mathrm{AB_{LOCC}}} }^{\dagger} E_{ ( \vec{s}_n ,\vec{t}_n  )}^{\mathrm{AB_{LOCC}}} = \Pi_1 ~~\mathrm{and}~~ \nonumber\\
&& \sum_{(\vec{v}_n,\vec{w}_n)}{E_{ ( \vec{v}_n ,\vec{w}_n  )}^{\mathrm{AB_{LOCC}}} }^{\dagger} {E_{ ( \vec{v}_n ,\vec{w}_n  )}^{\mathrm{AB_{LOCC}}}} =  \Pi_2  \nonumber
\eea
This shows that the LOCC protocol implements the corresponding POVM: $\{ \Pi_1, \Pi_2\}$, and hence, the LOCC protocol is optimal to discriminate between states $\{q_i,\rho_i \}_{i=1}^{2}$. 

$(\Rightarrow)$ Conversely, we assume that Alice and Bob can implement the optimal discrimination for states $\{q_i, \rho_i \}_{i=1}^2$ by an LOCC protocol. This implies that all $(\vec{k}_n,\vec{l}_n)$ can be partitioned into two classes $\{(\vec{s}_n,\vec{t}_n) \}$ and $\{(\vec{v}_n,\vec{w}_n)\}$ such that the given states are optimally discriminated by the POVM elements in the following
\bea
&& \Pi'_1 \equiv \sum_{(\vec{s}_n,\vec{t}_n)}{E_{ ( \vec{s}_n ,\vec{t}_n  )}^{\mathrm{AB_{LOCC}}} }^{\dagger} E_{ ( \vec{s}_n ,\vec{t}_n  )}^{\mathrm{AB_{LOCC}}} ~~\mathrm{and}~~ \nonumber \\
&& \Pi'_2 \equiv \sum_{(\vec{v}_n,\vec{w}_n)}{E_{ ( \vec{v}_n ,\vec{w}_n  )}^{\mathrm{AB_{LOCC}}} }^{\dagger} E_{ ( \vec{v}_n ,\vec{w}_n  )}^{\mathrm{AB_{LOCC}}}. \nonumber
\eea
In Ref. ~\cite{Helstrom:1969aa} it is shown that for two-state discrimination, POVM elements are unique, by which we have that $\Pi'_1= \Pi_1$ and $\Pi'_2 = \Pi_2$. Now note that the POVM elements $\Pi_1$ and $\Pi_2$, for two state discrimination, are projectors, hence $\Pi_1 \Pi_2 = 0$. This immediately implies that the states $\widetilde{\Pi}_1$ and $\widetilde{\Pi}_2$ can be perfectly discriminated by the same LOCC protocol.
\end{subequations}
\end{proof}

\subsubsection*{   LOCC measurements for optimal channel discrimination}

It is left to show that two states from an optimal measurement in Eq. (\ref{eq:two}),
\bea
\widetilde{\Pi}_1 = \Pi_1~~\mathrm{and}~~ \widetilde{\Pi}_2 =\Pi_2/3 \label{eq:twopi}
\eea
can be perfectly discriminated by LOCC. This implies that a pair of quantum states for which an optimal measurement is given by the POVM elements are optimally discriminated by LOCC. The necessary and sufficient condition for the perfect discrimination by LOCC is shown as follows. \\

{\bf Lemma}. The states $\widetilde{\Pi}_1$ and $\widetilde{\Pi}_2$ in Eq. (\ref{eq:twopi}) can be perfectly discriminated by LOCC if and only if $U \ket{\psi}$ is a product state. The perfect discrimination can be obtained by a one-way LOCC protocol. Moreover, for a two-qubit gate $U$ there exists a product state $\ket{\psi}$ such that the resulting state $U|\psi\rangle $ is a product state.
\\

A part of the proof can be found in Ref. \cite{6687245}, see also \cite{Owari_2008}. An alternative one is shown as follows. We first show that the states $\widetilde{\Pi}_1$ and $\widetilde{\Pi}_2$ can be perfectly discriminated by LOCC if and only if $U \ket{\psi}$ is a product state. A one-way LOCC protocol for the perfect discrimination is also provided. 

\begin{proof}

($\Leftarrow$) Suppose that $U \ket{\psi}$ be a product state, denoted by $U \ket{\psi}= \ket{c}\ket{d},$ where $\ket{c} \in \mathcal{H}_A$ and $\ket{d} \in \mathcal{H}_B$. Then, the other POVM has a decomposition as follows, 
\begin{align}
\label{eq:theory:thm1:1}
\widetilde{\Pi}_2 = \frac{1}{3} \left( \proj{c_\perp,d} + \proj{c,d_\perp} + \proj{c_\perp,d_\perp} \right), 
\end{align}
where $\braket{c}{c_\perp} = \braket{d}{d_\perp} = 0$. The LOCC protocol for perfect discrimination between $\widetilde{\Pi}_1$ and $\widetilde{\Pi}_2$ is straightforward. Alice applies measurement in the orthonormal basis $\{ \ket{c}, \ket{c_\perp} \}$ and Bob does also in the orthonormal basis $\{ \ket{d}, \ket{d_\perp} \}$. Then if Alice obtains the outcome $\ket{c}$ and Bob the outcome $\ket{d}$, they declare that state $\widetilde{\Pi}_1$ is shared. Otherwise, they conclude state $\widetilde{\Pi}_2$. In this way, two parties can perfectly discriminated between two state $\widetilde{\Pi}_1$ and $\widetilde{\Pi}_2$. \smallbreak

($\Rightarrow$) Conversely, suppose that states $\widetilde{\Pi}_1$ and $\widetilde{\Pi}_2$ can be perfectly discriminated by a one-way LOCC protocol. Let Alice start the protocol, and $K_A$ denotes one of the Kraus operators of Alice's measurement in the one-way protocol for perfect discrimination. Consequently, the post-measurement states are given by,
\bea
(K_A\otimes \mathrm{I})\widetilde{\Pi}_1 (K_A\otimes \mathrm{I})^{\dagger}~~\mathrm{and}~~(K_A\otimes \mathrm{I})\widetilde{\Pi}_2 (K_A\otimes \mathrm{I})^{\dagger}. ~~~~ \label{eq:cont}
\eea
Since a Kraus operator $K_A$ on Alice's side leads to perfect discrimination, the post-measurement states in the above are orthogonal, i.e.,
\bea
\tr[(K_A\otimes \mathrm{I})\widetilde{\Pi}_1 (K_A\otimes \mathrm{I})^{\dagger} (K_A\otimes \mathrm{I})\widetilde{\Pi}_2 (K_A\otimes \mathrm{I})^{\dagger}] =0\nonumber
\eea
Rewriting the equation in the above, one can find that $  \widetilde{\Pi}_1  (K_{A}^{\dagger} K_A \otimes \mathrm{I}) \widetilde{\Pi}_2=0$. Let $\{  |\phi_1\rangle, |\phi_2\rangle, |\phi_3\rangle  \}$ be an orthonormal basis for the support of $\widetilde{\Pi}_2$. It follows that,
\bea
\tr_A [ K_{A}^{\dagger} K_A ( \tr_B [  |\phi_j\rangle\langle \psi| U^{\dagger} ] ) ]  =0,~~\forall j\in \{1,2,3\}. \label{eq:list}
\eea
That is, measurement ${K_A}^\dag K_A $ is orthogonal to the reduced operator $\tr_B [  |\phi_j\rangle\langle \psi| U^{\dagger}]$ for all $j=1,2,3$. Let $U \ket{\psi} $ have the following Schmidt decomposition.
\bea
U \ket{\psi} = \mu \ket{c}\ket{d} + \sqrt{1 - \mu^2} \ket{c_\perp}\ket{d_\perp} \label{eq:assume}
\eea
Suppose $\mu \in (0,1)$ for which the state $U | \psi \rangle$ in the above is entangled. Since $\widetilde{\Pi}_1$ and $\widetilde{\Pi}_2$ are orthogonal, one can find that the states $\{|\phi_{j}\rangle \}_{j=1}^3$ in the support of $\widetilde{\Pi}_2$ are written as follows,
\bea
&&|\phi_1 \rangle = - \sqrt{1-\mu^2} |c\rangle |d\rangle + \mu \ket{c_\perp} \ket{d_\perp},~ \ket{\phi_2} =  \ket{c} \ket{d_\perp},~  \nonumber \\
&& \mathrm{and}~~~\ket{\phi_3} =\ket{c_\perp}\ket{d} \nonumber 
\eea
for $|c\rangle \in\H_A$ and $|d\rangle \in \H_B$ under the assumption that $\mu\in (0,1)$. Consequently, from Eq. (\ref{eq:list}) the resulting operators $ \tr_B \left(\ketbra{\phi_j}{\psi}U^\dag  \right)$ for $j=1,2,3$ on the Alice side, that are orthogonal to $ K_{A}^{\dagger} K_A $, can be obtained as follows,
\bea
&& \mu \sqrt{1-\mu^2} \left( \proj{c_\perp} - \proj{c} \right),\nonumber \\
 && \sqrt{1-\mu^2} \ketbra{c_\perp}{c}, ~\mathrm{and}~ \mu \ketbra{c}{c_\perp}. \label{eq:condition}
\eea
Then, we have $K_{A}^{\dagger} K_A \propto | c\rangle \langle c| + |c_{\perp}\rangle \langle c_{\perp}| =  \mathrm{I}$, that is, the measurement corresponds to an identity $\mathrm{I}$. This leads to the contradiction to the assumption that Alice's measurement can make two states in Eq. (\ref{eq:cont}) perfectly distinguishable, since the measurement is given by $\kappa\mathrm{I}$ for some $\kappa\in (0,1]$. Therefore, the state in Eq. (\ref{eq:assume}) is not entangled, i.e., we have $\mu =0$ or $\mu=1$, so that Alice's measurement $K_{A}^{\dagger} K_A$ can lead to perfect discrimination. We have shown that the state in Eq. (\ref{eq:assume}) is a product state. 
\end{proof}

\subsubsection*{   LOCC discrimination for two-qubit gates}

Then, for a two-qubit gate $U$, one can always find a product state $\ket{\psi}$ such that the resulting state $U|\psi\rangle $ is a product state. A two-qubit gate has a canonical decomposition as follows ~\cite{PhysRevA.63.062309, KHANEJA200111}, 
\begin{equation}
\label{U}
U = \left(  U_A \otimes U_B \right) U_d \left(V_A \otimes V_B \right),
\end{equation}
with $U_A$, $V_A$, $U_B$, and $V_B$ local unitary transformations and $U_d$ an entangling unitary transformation, 
\bea
&& U_d = \sum_{j=1}^{4} e^{i \lambda_j} \proj{\Phi_j} ~~ \mathrm{where} \label{eq:dunitary} \\
&& \ket{\Phi_1} = \dfrac{ 1}{\sqrt{2}} (\ket{00} + \ket{11} ),~ \ket{\Phi_2} =  \dfrac{1 }{\sqrt{2}} ( \ket{00} - \ket{11}), \nonumber \\
&&      \ket{\Phi_3} =  \dfrac{1 }{\sqrt{2}} ( \ket{01} - \ket{10}),~ \ket{\Phi_4} =  \dfrac{1}{\sqrt{2}} ( \ket{01} + \ket{10} ). \nonumber
\eea
In the following, we show that one can find a product state that is also a product state after an entangling gate \cite{Bae:2015aa}. Then, we extend it to arbitrary two-qubit gates. 

A two-qubit state can be written in the basis in the above, 
\bea
\ket{\psi} = \sum_{j=1}^{4} \alpha_j \ket{\Phi_j} \nonumber
\eea
which is a product state if and only if 
\bea
{\alpha}_1^2 - {\alpha}_2^2 + {\alpha}_3^2 - {\alpha}_4^2 = 0. \label{eq:p}
\eea 
After applying an entangling gate in Eq. (\ref{eq:dunitary}), the resulting state $U|\psi\rangle $ is separable if and only if 
\bea
\left(e^{ i \lambda_1} {\alpha}_1\right)^2 - \left(e^{ i \lambda_2} {\alpha}_2\right)^2 + \left(e^{ i \lambda_3} {\alpha}_3\right)^2 - \left(e^{ i \lambda_4} {\alpha}_4\right)^2 = 0 ~~~~~~~\label{eq:up}
\eea
One aims to find a vector $\vec{ \alpha } = (\alpha_{1}^2, \alpha_{2}^2, \alpha_{3}^2, \alpha_{4}^2)$ such that the conditions of a product state in Eqs. (\ref{eq:p}) and (\ref{eq:up}) are satisfied. Let us rewrite Eqs. (\ref{eq:p}) and (\ref{eq:up}) as conditions as follows, 
\bea
\vec{t} \cdot \vec{\alpha}  = 0,~ \vec{ u_{re}} \cdot \vec{\alpha} = 0, ~\mathrm{and}~\vec{ u_{im}} \cdot \vec{\alpha} =0 \label{eq:ortho}
\eea
where three vectors are defined as 
\bea
\vec{t} & = & (1,-1,1,-1)^T \nonumber \\
u_{re} & = & (  \cos 2\lambda_1 , - \cos 2\lambda_2, \cos 2\lambda_3, - \cos 2\lambda_4)^T \nonumber \\
u_{im} & = & (  \sin 2\lambda_1 , - \sin 2\lambda_2, \sin 2\lambda_3, - \sin 2\lambda_4)^T. \nonumber 
\eea
These vectors define a subspace, denoted by $S_{U}$ {  in} $\mathbb{R}^4$, whose dimension is less than or equal to three, i.e., $\mathrm{dim} S_U  \le  3$. Hence, its orthogonal complement subspace $S_{U}^{\perp}$ is not a null space, i.e., $\mathrm{dim} S_{U}^{\perp}\ge 1$. Since $\vec{\alpha}$ is orthogonal to the subspace $S_{U}$, one can always find $\vec{\alpha} \in S_{U}^{\perp}$ that satisfies the conditions in Eq. (\ref{eq:ortho}). This shows the existence of a product state $|\psi\rangle$ that remains separable after application of an entangling unitary transformation. 

The result can be extended to two-qubit gates in general. From the results shown so far, for an entangling unitary gate $U_d$ one can always find a product state $\ket{\psi}=  |a\rangle |b \rangle$ for some $|a\rangle\in\H_A$ and $|b\rangle \in \H_B$ such that such that $U_d \ket{a}\ket{b}$ is a product state, denoted by  $U_d \ket{a}\ket{b}= \ket{c}\ket{d}$. For an arbitrary two-qubit gate in Eq. (\ref{U}), one can choose 
\bea
\ket{\psi} = \left(V_A^\dag \otimes V_B^\dag \right) \ket{a} \ket{b} \nonumber
\eea
so that $U \ket{\psi} = \left( U_A \otimes U_B \right) \ket{c}\ket{d}$, which is also a product state.  

It is thus shown that for a two-qubit gate $U$ and its noisy counterparts $[\N_{U}^{p}]$, an LOCC discrimination can achieve the guessing probability in Eq. (\ref{eq:pg2}): there exists a product state $|\psi\rangle$ that leads to $U|\psi\rangle$, also a product state, by which the resulting states $U|\psi\rangle$ and $\N_{U}^{p}[|\psi\rangle]$ are optimally discriminated by an LOCC protocol.  

\subsection{  LOCC protocols for channel discrimination} 

An LOCC protocol to optimally discriminate between a two-qubit gate $U$ and $\N_{U}^{p}$ works as follows. For convenience, let Alice and Bob hold single qubits, respectively, in a product state $|\psi\rangle$ such that $U|\psi\rangle $ is a product state. An optimal measurement can be written in a decomposition as follows,
\bea
\Pi_1  & = &   (U_A \otimes U_B) | c,d\rangle \langle c,d | (U_{A} \otimes U_B)^{\dagger} \nonumber \\
\Pi_2  & = &    \left( U_A \otimes U_B \right)  ( \mathrm{I}_A \otimes \mathrm{I}_B - | c,d\rangle \langle c,d |)  \left(U_A \otimes U_B\right)^\dag \nonumber 
\eea
for some single-qubit unitaries $U_A$ and $U_B$ and orthonormal basis $\{ |c\rangle, |c_{\perp}\rangle\}$ and $\{ |d \rangle, |d_{\perp}\rangle\}$. Then, Alice and Bob perform measurements $\left\{ U_A\ket{c}, U_A \ket{c_\perp} \right\}$ and $\left\{ U_B\ket{d}, U_B \ket{d_\perp} \right\}$, respectively, and they communicate the measurement outcomes. When the outcome is found as $U_A\ket{c}$ and $U_B\ket{d}$, they conclude that a state $\widetilde{\Pi}_1$ is shared, and consequently a two-qubit gate $U$. For other outcomes, they conclude the state $\widetilde{\Pi}_2$, and consequently a noisy one $\N_{U}^{p}$, i.e., the presence of depolarization noise. The probability of a correct conclusion is given by the guessing probability in Eq. (\ref{eq:pg2}).

We have thus devised a scheme of a single-copy certification of a two-qubit gate in the presence of depolarization noise. It is cost efficient: a realization of a two-qubit gate is certified by a single-shot measurement without entangled resources. As an instance, one may consider a complete depolarization noise $D$ that appears with the {\it a priori} probability $q=1/2$. The presented scheme certifies a realization of a two-qubit gate $U$ by a single-shot measurement, where the probability of making a correct conclusion is given as $7/8=0.875$. 
\\

\subsubsection*{ Example: CNOT gate}


We here consider an example with a CNOT gate, that plays a significant role in quantum information processing in general as it generates entangled states and composes a set of universal gates. In fact, concatenation of CNOT gates with a sufficiently high precision is a key to universal quantum computation. Then, a single-shot certification for a realization of a CNOT gate enables an experimentalist to efficiently conclude if noise is present in a realization. 

As we have explained above, a state $\ket{\psi}= \ket{0}\ket{0}$ is chosen such that it remains as a product state under a CNOT gate, i.e. $U |00\rangle = |00\rangle$. POVM elements for the optimal discrimination are obtained as follows,
\bea
{ \Pi_1} & = & |0  \rangle_{A } \langle 0 | \otimes |0  \rangle_{B } \langle 0 | \nonumber  \\
{ \Pi_2} & = &  \mathrm{I}_A \otimes \mathrm{I}_B -  |0  \rangle_{A } \langle 0 | \otimes |0  \rangle_{B } \langle 0 |. \label{eq:optcnot} 
\eea
Then, Alice and Bob perform measurements $\{ \ket{0}_{A}, \ket{1}_{A} \}$ and $\{ \ket{0}_{B}, \ket{1}_{B} \}$ and communicate their measurement outcome. If they find the outcome $00$, they conclude the gate operation is noiseless, i.e., $U$. Otherwise, for outcomes $01$, $10$, and $11$, they conclude that depolarization noise is present.  

In the example above, the guessing probability in Eq. (\ref{eq:pg2}) can be reproduced as follows. For the input state $|\psi\rangle$, suppose that a CNOT gate is applied with {\it a priori} probability $1-q$. In this case, the outcome must be $00$. Or, if its noisy counterpart $\N_{U}^{p}$ is applied  {\it a priori} probability $q$, the outcome would be $00$ with a probability $3p/4$ or $01$, $10$ and $11$ with an equal probability $p/4$ respectively. Thus, the probability of making a correct guess is as follows, 
\bea
p_{\g}^{(\mathrm{LOCC})} = (1-q)\times 1 + q\times 3\times \frac{p}{4} \label{eq:ex}
\eea
which is equal to the guessing probability in Eq. (\ref{eq:pg2}). We recall that Eq. (\ref{eq:ex}) is valid when $1-2q +\frac{3}{4}pq\geq 0$, see Eq. (\ref{eq:pg2}). Note that an input state can be chosen by $| i_0 j_0 \rangle$ for $i_0,j_0=0,1$. For instance, for an input state $|11\rangle$, the outcome $10$ leads to the conclusion of a CNOT gate and other outcomes to its noisy counterpart.  


\section{Experiment}

We then apply LOCC protocols devised in the previous section to an experimental realization of a two-qubit gate for the single-shot certification of the implementation. Note that the scheme can efficiently certify a realization of two-qubit gate with minimal resources. 

\begin{figure}[t]
\begin{center}
\includegraphics[width= 3.4in]{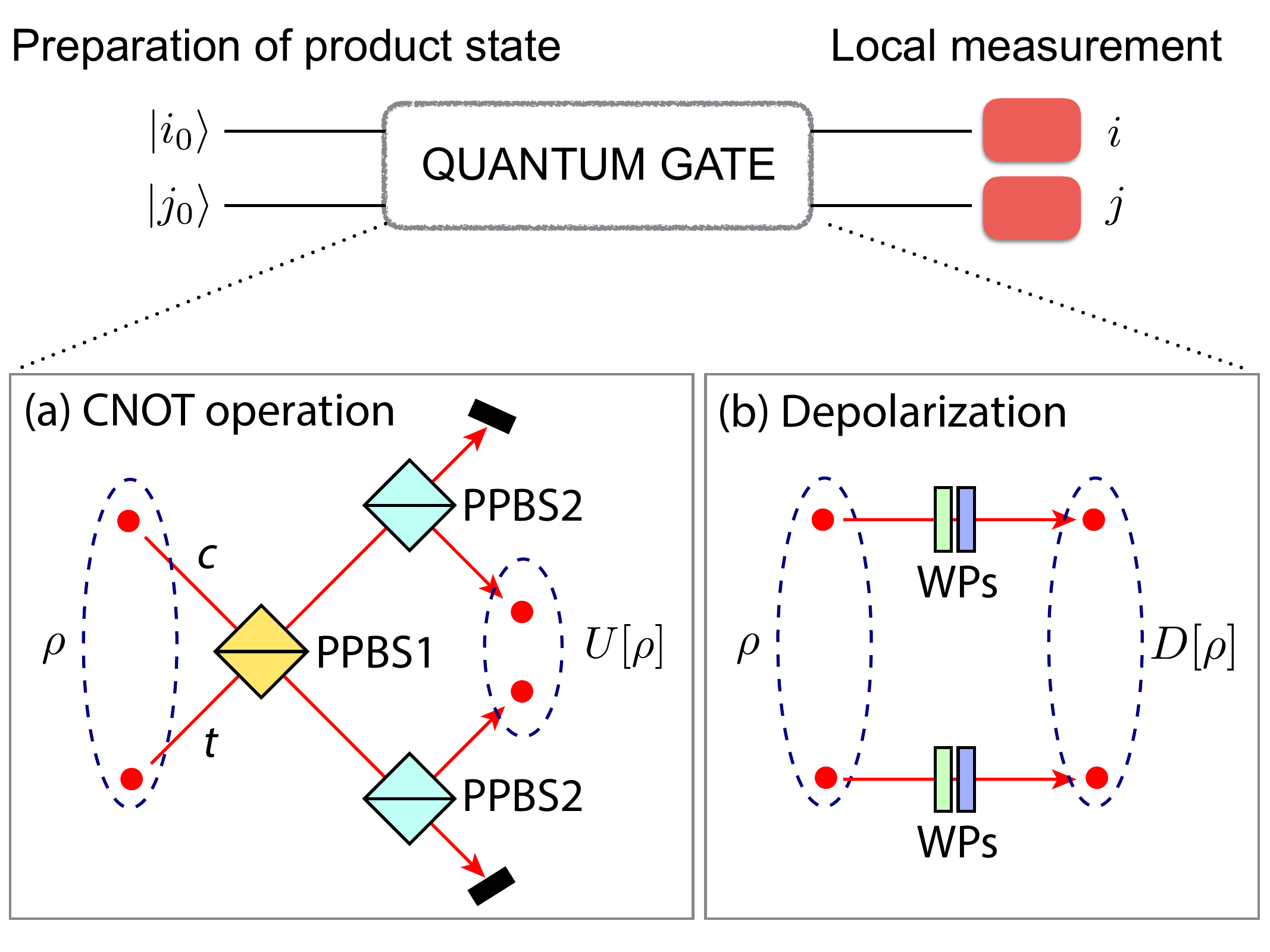}
\caption{ (a) A CNOT gate is implemented by three partial polarizing beam splitters (PPBS) with $T_H=1$ and $T_V=1/3$ (PPBS1) and $T_H=2/3$ and $T_V=0$ (PPBS2). (b) Depolarization noise for two-qubit states is realized by averaging $16$ combinations of Pauli operations, which are implemented by waveplates, with equal probabilities. A single-copy certification is demonstrated with a product state and an individual measurement.  }
\label{setup}
\end{center}
\end{figure}

 In particular, a CNOT gate for photonic qubits are considered, for which the certification scheme shown in the example above is applied. For photonic qubits, a CNOT gate proposed in Refs. \cite{PhysRevLett.95.210504, PhysRevLett.95.210505, PhysRevLett.95.210506, Crespi:2011aa}, see also Fig.~\ref{setup}(a), is realized and certified. Then, the LOCC protocol devised in the previous subsection is applied to find if depolarization noise is present. The experimental details of the single photon preparation and the noise-adjustable CNOT gate implementation are presented as follows. 

\subsection{Single-photon source}

Single-photon states are generated by Type-\Romannum{1} spontaneous parametric down-conversion at a 6~mm $\beta$-Barium borate crystal pumped by a 408~nm diode laser~\cite{PhysRevA.86.042334}. The spectrum of the single-photon pair is chosen by interference filters whose central wavelength is 816 nm and full width at half maximum is 5 nm.

\subsection{  Implementation of noise-adjustable CNOT gate}
 
The noisy CNOT operation $\N_{U}^{p}$ has a form of the incoherent mixture of a CNOT and the depolarizing channels, and thus one can implement it by mixing the results of CNOT and depolarizing operations. The depolarizing noise probability $p$ can be adjusted by changing the ratio between the CNOT and the depolarizing operations.

The experimental setup of realizing a CNOT gate for photonic polarization qubits is as follows, see also Fig. \ref{setup}. By defining the computational basis $\{|0\rangle,|1\rangle \}$ for control $(c)$ and target $(t)$ qubits as $\{|H\rangle,|V\rangle\}$, and $\{|D\rangle,|A\rangle\}$, respectively, a photonic CNOT gate can be implemented with three partial-polarizing beam splitters (PPBS) \cite{PhysRevLett.95.210504, PhysRevLett.95.210505, PhysRevLett.95.210506, Crespi:2011aa}. Here, $|H\rangle$, $|V\rangle$, $|D\rangle$, and $|A\rangle$ refer horizontal, vertical, diagonal, and anti-diagonal polarization states, respectively. The transmissivities of PPBS1 are $T_H=1$ for horizontal polarization and $T_V=1/3$ for vertical polarization, while those of PPBS2 are $T_H=2/3$, and $T_V=0$, respectively. Note that the success operation of CNOT corresponds to the case when a single-photon is found at each output and the success probability is $1/9$. 

The experimental realization of a CNOT gate can be assessed by the quantum process fidelity. Let $\chi_{0}$ and $\chi_{\mathrm{ex}}$ the Choi-Jamilokowski operators of a CNOT gate $U$ and its experimental realization $\N_{\mathrm{ex}} $, respectively. The operator $\chi_{\mathrm{ex}}$ can be constructed by quantum process tomography~\cite{PhysRevA.63.020101}. Then, the process fidelity for the quantification of $\N_{\mathrm{ex}}$ is given by $F_{\mathrm{process}} = \tr[\chi_0 \chi_{\mathrm{ex}}]$.

\begin{figure}[t]
\begin{center}
\includegraphics[width= 3.2in]{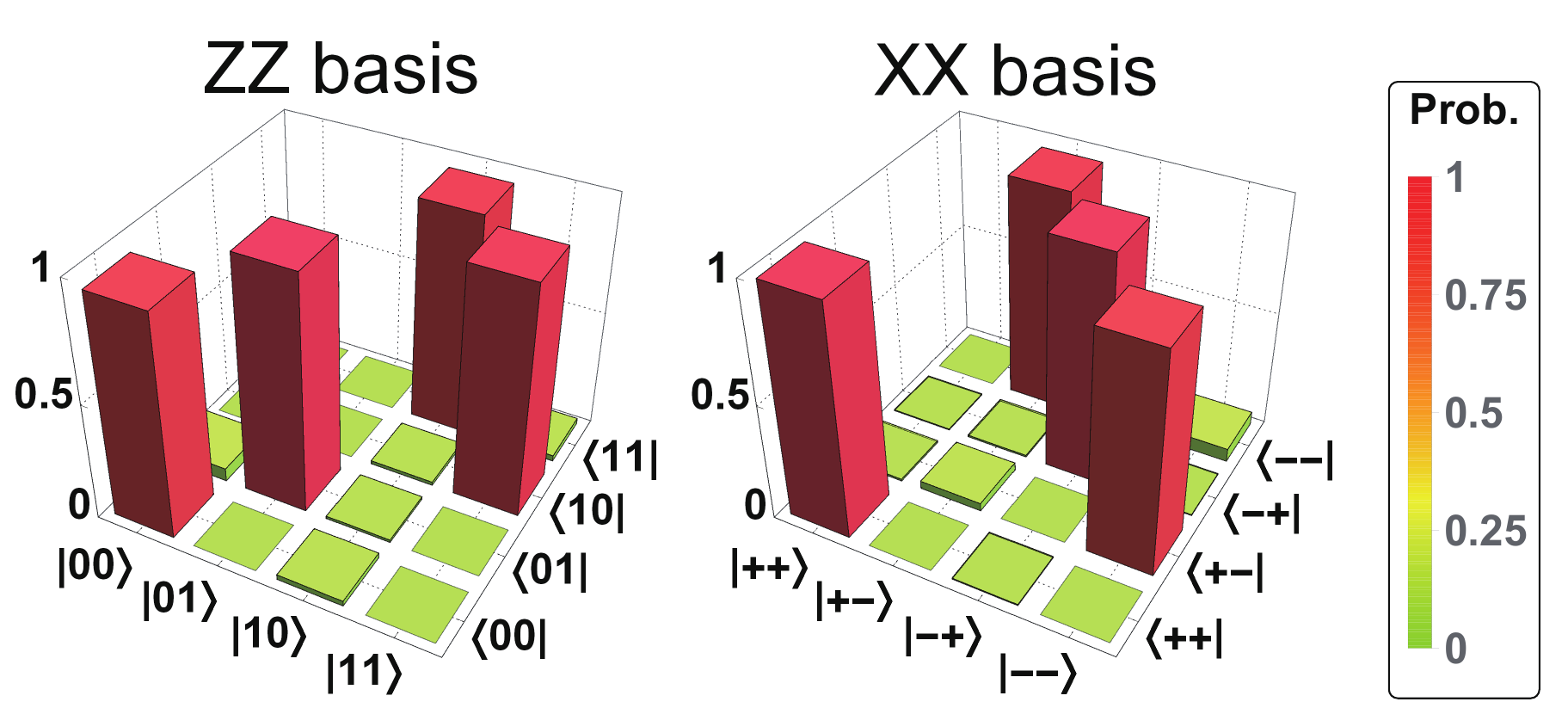}
\caption{Truth tables of CNOT gate from experimental data are shown in $ZZ$  and $XX$ bases. The average fidelities of truth tables are given as $0.96 \pm 0.01$ and $0.96 \pm 0.02$, for $ZZ$ and $XX$ bases, respectively.}\label{truth}
\end{center}
\end{figure}

In Fig.~\ref{truth}, the truth tables for $ZZ$ and $XX$ bases are shown from experimental data. Here, $ZZ~(XX)$ basis denotes that both input and output states are analyzed with $Z~(X)$ basis where $Z~(X)$ basis corresponds to $\{|0\rangle, |1\rangle\} (\{|+\rangle, |-\rangle\})$ where $|\pm\rangle=\frac{1}{\sqrt{2}}(|0\rangle\pm|1\rangle)$. The process fidelities for $ZZ$ and $XX$ bases are given as $F_{ZZ}=0.96\pm0.01$, and $F_{XX}=0.96\pm0.02$, respectively. From these, we have $F_{\rm process} \in [0.92, 0.96]$, which follows from the relation $F_{ZZ}+F_{XX}-1 \leq F_{\mathrm{process}} \leq \min(F_{ZZ},F_{XX})$~\cite{PhysRevLett.94.160504}. 
 
 
The depolarizing noise channel for two-qubit states can be implemented by averaging over all the 16 combinations of single-qubit Pauli operations with equal probability, i.e., $D[\rho] = \frac{1}{16}\sum_{i,j=0}^{3}\sigma_{i}\otimes \sigma_{j} \rho \sigma_{i} \otimes \sigma_{j}$ where $\{ \sigma_i\}_{i=0}^3$ denote Pauli matrices \cite{PhysRevLett.107.160401}. The single-qubit Pauli operations can be easily implemented by sets of waveplates.

\begin{figure}[t]
\includegraphics[width= 3.2in]{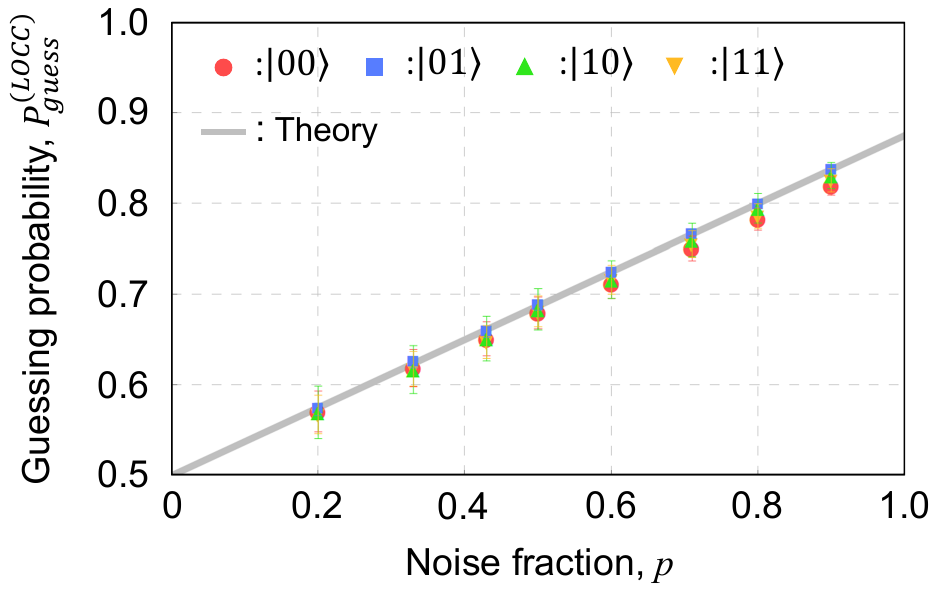}
\caption{The guessing probability in Eq. (\ref{eq:pexp}) for a CNOT gate and its noisy counterpart is shown in the case of equal {\it a priori} probabilities $q=1/2$. The solid line corresponds to the guessing probability in Eq. (\ref{eq:ex}). In experiment, different input states $|ij\rangle$ for $i,j=0,1$ are applied and shown with different colors. Error bars are experimentally obtained standard deviations.}
\label{guessing}
\end{figure}

\subsection{ Certification of a CNOT gate with LOCC} 


We now apply the single-shot certification scheme to a realization of a CNOT gate for photonic qubits, see also the subsection with an example of a CNOT gate. Let $\E \in \{ U, \N_{U}^{p}\}$ denote one of a CNOT gate or its noisy one for some $p$. Once it is chosen, the operation is applied to an input state $|00\rangle$ repeatedly. As it is shown in Eq. (\ref{eq:optcnot}), an individual measurement is performed in the basis $\{ |0\rangle_A,|1\rangle_A\}$ and $\{ |0\rangle_B,|1\rangle_B\}$.

To experimentally demonstrate the guessing probability, let $n_{ij} (\E)$ denote the coincidence counts in a measurement outcome $ij$, and the total counts is given by $n (\E)=\sum_{i,j=0,1} n_{ij} (\E)$. For outcomes $00$, a CNOT gate is concluded. For the others, its noisy counterpart is concluded. The probability that the strategy gives a correct conclusion is found as follows
\bea
P_{{\rm guess},est}^{\rm (LOCC)}=\frac{1}{2} \times \frac{n_{00} ( U )}{ n(U)  } + \frac{1}{2}\times \frac{\sum_{(i,j)\neq(0,0)} n_{ij}(  \N_{U}^{p} ) }{ n (  \N_{U}^{p} ) }. ~~~~\label{eq:pexp}
\eea
In Fig.~\ref{guessing}, the guessing probability in Eq.~(\ref{eq:pexp}) is shown, having a good agreement with Eq. (\ref{eq:ex}). For an input state $|11\rangle$, the measurement outcome $10$ concludes a CNOT gate and otherwise, a noisy one. 

\begin{figure}[t]
\includegraphics[width= 3.2in]{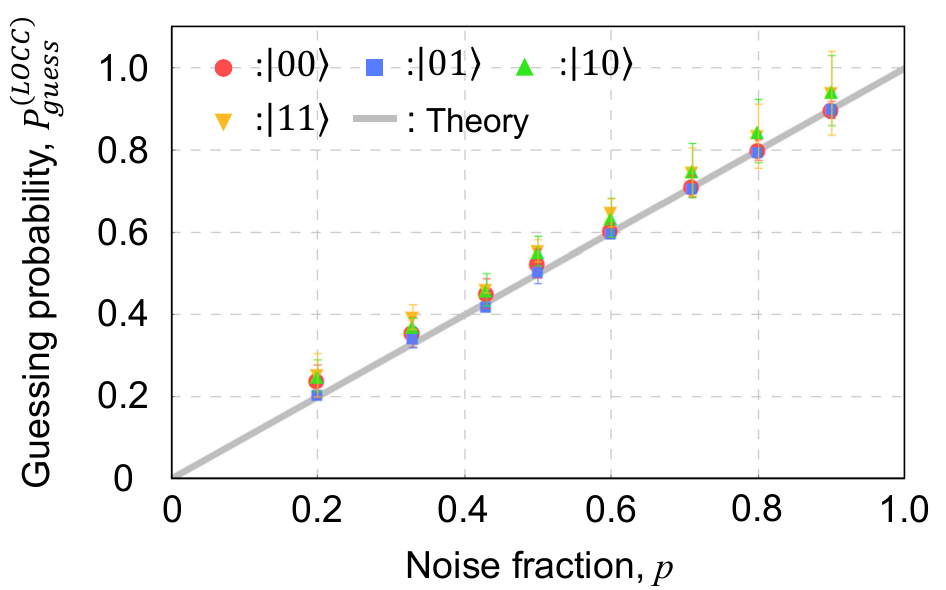}
\caption{ A noise fraction is estimated by repeating a measurement. In experiment, different input states $|ij\rangle$ for $i,j=0,1$ are applied. The experimental data with different input states are presented with different marks with different colors. Error bars are experimentally obtained standard deviations.}
\label{noise}
\end{figure}

As a byproduct, the presented LOCC scheme can be used to estimate a noise fraction $p$. Suppose that a noisy channel $\N_{U}^p$ for some unknown $p$ is applied all the time, i.e., $q=1$. Let $p_{ij}$ denote the probability of an outcome $ij$. It holds that $
p_{ij} = (1-p) ~p_{ij} (U)   + p ~p_{ij} (D)$ where $p_{ij} (\E)$ is the probability of an outcome $ij$ for an operation $\E \in \{ U, D\}$. For a CNOT gate, we have that $p_{00}(U)=1$ and $p_{ij} (D)=1/4$ for all $i,j=0,1$. This finds the noise fraction $p = 4 (1- p_{00})/3$. We have used $n_{ij} (\N_{U}^p)$ as the coincidence counts on a measurement with the computational basis $|ij\rangle$, from which a noise fraction estimated from experimental data can be found,
\begin{equation}
p_{est} = \frac{4}{3} \frac{  n(\N_{U}^{p}) - n_{00} (\N_{U}^{p})  }{   n (\N_{U}^{p}) }, \label{eq:estp}
\end{equation}
From experimental data, the noise fractions $p_{est}$ can be estimated as follows. If a measurement in the computational basis gives $00$, it is concluded that a CNOT gate has been applied. For other outcomes, $01$, $10$, and $11$, a noisy CNOT operation is concluded. The experimental proof-of-principle demonstration has considered other input states, $|01\rangle$, $|10\rangle$ and $|11\rangle$. In general, for input state $|ij\rangle$, the measurement outcome $i$ and $j\oplus i$ concludes a CNOT gate. The experimental results are shown in Fig. \ref{noise}. Otherwise, a noisy operation is concluded. It is worth mentioning the argument used for a CNOT gate also applies to a SWAP gate that works as $U \ket{a}_A\ket{b}_B = \ket{b}_A \ket{a}_B$.

\section{Conclusion}

In conclusion, we have shown an optimal single-copy certification for a two-qubit gate realized in a circuit in the presence of depolarization noise. Technically, the certification scheme corresponds to distinguishing a two-qubit gate from a set of two-qubit channels. Since an optimal measurement for the discrimination is identical for all channels in the set, it is possible to exploit an optimal channel discrimination for the certification of a two-qubit gate. It is worth to mention that the probability from optimal discrimination is limited by fundamental principles, see e.g. \cite{EPTCS171.3}. 

It is shown that a maximal probability in a single-copy certification can be achieved without an entangled resource at all. We emphasize that our scheme is thus feasible with NISQ technologies where single-qubit operations have much low error rates of order $0.1\%$. An experimental proof-of-principle demonstration is presented with photonic polarization qubits with LOCC only. A single-copy certification is demonstrated for a CNOT gate. The scheme is cost efficient, e.g., a realization of a CNOT gate in the presence of depolarization noise has been certified by the proposed single-shot and LOCC scheme. The certification scheme can be used to estimate a noise fraction existing in an experimental realization of a two-qubit gate. 

As a quantum circuit is composed of universal gates where a number of two-qubit gates are contained, our results opens a new avenue to efficiently certify two-qubit gates with minimal resources in a realistic scenario. For future directions, it would be interesting to extend to concatenated two-qubit gates and devise a single-shot certification for multiple gates. It is also interesting to consider the cases when more than single copies are available so that the guessing probability can be improved in a finite-copy scenario \cite{PhysRevLett.87.177901, PhysRevLett.99.170401, PhysRevLett.100.020503}. On a fundamental side, we leave it an open question when a measurement of an optimal discrimination of two channels can be extended to two sets of channels. 

 

\section*{Acknowledgement}
This work was supported by Samsung Research Funding \& Incubation Center of Samsung Electronics (Project No. SRFC-TF2003-01) and the KIST Institutional Programs (2E31021). 







%

\end{document}